\documentstyle[aps,prl,multicol,epsf]{revtex}

\begin{document}
\draft
\title{From mesoscopic magnetism
       to the anomalous 0.7 conductance plateau}
\author{Henrik Bruus, Vadim V.\ Cheianov, and Karsten Flensberg}
\address{{\O}rsted Laboratory, Niels Bohr Institute,
Universitetsparken 5, DK-2100 Copenhagen \O, Denmark}
\date{February 22, 2000}
\maketitle
\begin{abstract}
We present a simple phenomenological model which offers a 
unifying interpretation of the experimental observations on the
0.7 conductance anomaly of quantum point contacts. The model utilizes
the Landauer-B\"{u}ttiker formalism and involves enhanced spin
correlations and thermal depopulation of spin subbands. In particular
our model can account for the plateau value 0.7 and the unusual
temperature and magnetic field dependence. Furthermore it predicts an
anomalous suppression of shot noise at the 0.7 plateau. 
\end{abstract}
\begin{multicols}{2}

It has been known and well understood since
1988\cite{vanWees88,Wharam88,vanHouten92} that the dc-conductance $G$
of narrow quantum point contacts and quantum wires (both referred
to as QPCs below) is quantized in units of $G_2= 2\: e^2/h$. During
the past five years an increasing part of the experimental and
theoretical work on QPCs has been devoted to studies of deviations
from this integer quantization. In particular the discovery of the
0.7 conductance anomaly in 1996\cite{Thomas96} posed one of the
most intriguing and challenging puzzles in the
field\cite{Thomas98,Liang99,Kristensen98a,Kristensen98b,Kristensen98c,Reilly00,Berggren,Spivak}.
This anomaly is a narrow plateau, or in some cases just a
plateau-like feature appearing in scans of $G$ versus gate
voltage $V_g$ at a value of $G$ which is reduced by a factor 0.7 relative
to the ideal value $G_2$. The 0.7 conductance anomaly has been
recorded in in numerous QPC transport experiments (even before it
was noted in 1996, see {\it e.g.\/} Ref.~\cite{vanWees88})
involving many different materials, geometries and measurement
techniques. It can therefore be regarded as a universal effect.

Due to its universal character and the absence of a theoretical
understanding the 0.7 anomaly has been subject to intensive
experimental studies. In this paper we show that many of the
experimental findings can in fact be consistently interpreted
by invoking a model of enhanced spin correlations, both
spatially and temporally, of the charge carriers
in the QPC interaction. Due to the low density
the exchange interaction between electrons
in the QPC is strong and hence
there is a tendency to lign up the electron spins there.
Based on this physical picture, we formulate a simple
phenomenological model of tendency to form partially polarized states,
which together with the Landauer-B\"{u}ttiker (LB)
formalism naturally explains many experimental features of the
0.7 anomaly.

At this point it is important to mention that under very general
conditions truly 1d systems cannot exhibit ferromagnetic ordering
at all\cite{Lieb,Mermin}. Therefore we emphasize that the model we
are presenting does not rely on having a static magnetic moment,
but only of having a dynamical mesoscopic polarization, where the
correlation length is longer than the size of the QPC, and where the
correlation time is longer than the passage time through the
constriction.  

{\em Summary of experimental facts}.
Although the 0.7 anomaly has been observed in many other
experiments we refer mainly to the work of the Cambridge group
\cite{Thomas96,Thomas98,Liang99} and the Copenhagen group
\cite{Kristensen98a,Kristensen98b,Kristensen98c} presenting detailed
studies of the magnetic field and temperature dependence of the
anomaly. We emphasize that we are not dealing with the overall
suppression of the conductance plateaus which has been seen in some
samples\cite{Tarucha95,Yacobi96,Tscheuschner96}, and which
has been attributed to effects exterior to the contact
region\cite{Alekseev98}. 

The main experimental features of the 0.7 anomaly are:

(e1) The anomalous plateau is observed in a large variety of QPCs
at a value $G = \gamma\: G_2$, where the
suppression factor $\gamma$ is close to 0.7
\cite{Thomas96,Thomas98,Liang99,Kristensen98a,Kristensen98b,Kristensen98c}.
A typical semiconductor QPC has a width less than
0.1~$\mu$m and a length in the range 0.1 to 10~$\mu$m.

(e2) The temperature dependence is qualitatively the same for all
samples: the anomalous plateau is fully developed in some (device
dependent) temperature range typically above 2~K. With increasing
temperature both the anomalous and the integer plateaus vanish by
thermal smearing, while with decreasing temperature the width of the
anomalous plateau shrinks and the value of the suppression factor
$\gamma$ approaches 1
\cite{Thomas96,Thomas98,Liang99,Kristensen98a,Kristensen98b,Kristensen98c}.

(e3) A detailed study of the temperature dependence of $\gamma$ in QPCs
with a particularly large subband separation shows that in the
low temperature regime the conductance suppression has an
activated behavior: $1 - \gamma(T) \propto \exp(-T_a/T)$
\cite{Kristensen98b,Kristensen98c}.

(e4) The activation temperature $T_a$ is a function of 
$V_g$ vanishing at some critical gate voltage $V_g^0$. Close to
$V_g^0$ the dependence of $T_a$ on $V_g$ is well approximated by a
power law\cite{Kristensen98b,Kristensen98c},
$T_a \propto (V_g - V_g^0)^{\alpha}$, with $\alpha \approx 2$.

(e5) At a fixed temperature corresponding to a well developed 0.7
plateau, $\gamma$ shows a strong dependence on an
in-plane magnetic field\cite{Thomas96,Thomas98,Liang99}.
With increasing magnetic field $\gamma$ smoothly decreases from 0.7
at $B=0$~T to 0.5 at $B=13$~T. The latter value corresponds to the
expected LB conductance of one spin split subband.

(e6) Under the same temperature conditions as in (e5)
the 0.7 anomaly depends on the source-drain
bias. The suppression factor $\gamma$ increases smoothly from $\sim 0.7$
at zero bias to $\sim 0.9$ at large bias ($\sim 2$~mV)
\cite{Kristensen98c}.

{\em Alternative explanations for the 0.7 anomaly}. First we can
rule out impurity backscattering for two reasons (1) it
would lead to a non-universal suppression of conductance with a
strong sample dependent dependence on $V_g$ and (2) the
temperature dependence expected from thermal smearing of the LB
conductance is found to be much weaker than the observed
dependence of the 0.7 anomaly \cite{Kristensen98b,Kristensen98c}.
Thus a single particle picture cannot
explain the effect. With inclusion of electron-electron
interactions, a strong temperature dependence of conductance
suppression has been shown to arise due to an interaction induced
renormalization of backscattering, but the temperature
dependence is opposite to the observed
one\cite{Kane92,Tarucha95,Maslov,Maslov95}. We also note that in
the framework of Luttinger liquid theory interaction effects alone
has no effect on the
dc-conductance\cite{Maslov95,Ponomarenko95,Safi95}. Mechanism
based on activated backscattering, has also been
suggested\cite{Bruus}, but like for the impurity
backscattering suggestion, such a model cannot possibly offer an
explanation neither for the existence of the plateau nor for its
value.

Already in the first paper\cite{Thomas96} it was pointed out that
due to its magnetic field dependence the 0.7 anomaly is related to
spin polarization. This idea has been elaborated on in theoretical
papers\cite{Berggren,Spivak}, however, none of these approaches
have explained all of the experimental facts, and most strikingly
they predict plateaus at $G = G_2$ or $0.5\: G_2$ instead of as
the observed $0.7\: G_2$.

\begin{figure}
\centerline{\epsfysize=35mm\epsfbox{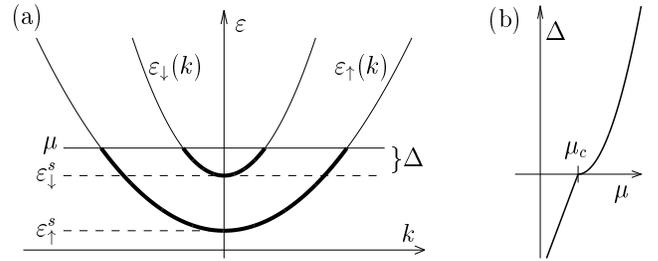}}
\narrowtext
\caption{ \label{fig:model}
(a) The instantaneous spin split subband structure of our model. (b)
The functional form Eq.~(\ref{eq:Deltamu}) of 
$\Delta(\mu) = \mu-\varepsilon^s_\downarrow(\mu)$ leading to the
anomalous 0.7 plateau.
}
\end{figure}

{\em The phenomenological model}.    
In our model we assume that the transmission coefficient
of electrons can be calculated in a ``frozen" configuration of
spin in the mesoscopic constriction. The dynamics of collective
degrees of freedom describing 
fluctuations of spin, are assumed to happen on larger time-scales.
Also the distribution of spin is assumed to be smooth such that an 
adiabatic approximation is valid. In the ``frozen spin configuration"
the transmission coefficient ${\cal T}^{\rm tot}_{\sigma}$ 
for a spin-$\sigma$ electron going through the QPC
can thus be calculated as 
${\cal T}^{\rm tot}_{\sigma} = {\cal T}_{\sigma}(E)
P_{\sigma} + {\cal T}_{\bar{\sigma}}(E) P_{\bar{\sigma}}$. Here
$P_{\sigma}$ ($P_{\bar{\sigma}}$) is the probability of finding
the incoming spin parallel (antiparallel) to the instantaneous
polarization. In
the isotropic case with $P_{\sigma} = P_{\bar{\sigma}}$ this leads
to the same results as a static situation where two spin subbands
are formed as shown in Fig.~\ref{fig:model}a, and therefore for 
simplicity we adopt this picture in the following modeling. 
Let the energy dispersion laws be given as

\begin{equation} \label{eq:dispersion}
\varepsilon_{\sigma}(k) =
\varepsilon_{\sigma}^0(k) + \varepsilon^s_{\sigma}, \quad
\sigma = \downarrow,\uparrow,
\end{equation}
where $\varepsilon_{\sigma}^0(k) \rightarrow 0$ for $k \rightarrow
0$ and $\varepsilon_{\sigma}^s$ is the subband edge. 
The system is partially polarized if the chemical potential
$\mu$ and the subband edges satisfy $\varepsilon^s_{\uparrow}(\mu)
< \varepsilon^s_{\downarrow}(\mu) < \mu$, where we have explicitly
indicated the $\mu$-dependence of the subband edges. Given this
model, at finite temperature $T$ using an idealized step-function 
transmission coefficient the LB conductance $G(T)$ of this system
is\cite{vanHouten92}
\begin{equation} \label{eq:dIdmu}
G(T)  = \frac{1}{2} G_2
\sum_{\sigma=\uparrow,\downarrow}
\int_{-\infty}^{\infty} d\varepsilon \:
\Theta(\varepsilon-\varepsilon^s_\sigma) \: f'[\varepsilon-\mu],
\end{equation}
where $f'$ is the derivative of the Fermi-Dirac distribution
$f[x]=[\exp(x/k_BT)+1]^{-1}$ and $\Theta(x)$ is the step function.
By integration we obtain
\begin{equation} \label{eq:conductance}
G(T)= \frac{1}{2} G_2(
f[\varepsilon^s_\uparrow(\mu)-\mu] +
f[\varepsilon^s_\downarrow(\mu)-\mu]).
\end{equation}
The important parameter is the spin down Fermi energy $\Delta(\mu)$
given by the energy difference between $\mu$ and
the minority spin subband edge (see Fig.~\ref{fig:model}):  
\begin{equation} \label{eq:delta}
\Delta(\mu) = \mu-\varepsilon^s_\downarrow(\mu).
\end{equation}
Consider now the situation where the spin polarization is nearly
complete, {\it i.e.\/} $\Delta(\mu) \ll
\varepsilon^s_\downarrow(\mu)-\varepsilon^s_\uparrow(\mu)$. In this
case three distinct temperature regimes exist.
In the high temperature regime, $k_B T \gg
\varepsilon^s_\downarrow(\mu)-\varepsilon^s_\uparrow(\mu)$,
both terms in Eq.~(\ref{eq:conductance}) are 0.5 so that $G = 0.5\: G_2$.
At low temperatures, $k_B T \ll \Delta(\mu)$,
both terms in Eq.~(\ref{eq:conductance}) are 1 and the conductance is
the usual $G_2$. Remarkably, in the entire temperature range

\begin{equation} \label{eq:condition}
\Delta(\mu) \ll k_B T \ll
\varepsilon^s_\downarrow(\mu)-\varepsilon^s_\uparrow(\mu),
\end{equation}
the contribution of the first term is 0.5 while the second term
remains 1 yielding $G = 0.75\: G_2$, and the magic number
$\approx 0.7$ emerges. Thus a 0.7 quasi-plateau appears if the
condition (\ref{eq:condition}) is fulfilled for a sufficiently broad
range of $\mu$ (in experiments $\mu \propto V_g$). In fact, below we
argue that it follows from general considerations that the functional
form of $\Delta(\mu)$, also shown in Fig.~1(b), is
\begin{equation}  \label{eq:Deltamu}
\Delta (\mu )= \left\{
\begin{array}{ll}
C(\mu -\mu _{c})^{2},& {\rm for\:} \mu > \mu_c\\
D(\mu -\mu _{c}),& {\rm for\:} \mu < \mu_c
\end{array} \right.
\end{equation}
which exactly expresses the tendency for
$\varepsilon_{\downarrow}^{b}(\mu)$ to lock onto the value of
$\mu$ by keeping $\Delta(\mu)$ small. We derive
Eq.~(\ref{eq:Deltamu}) starting from a local spin density functional:
$F=E[n_{\downarrow},n_{\uparrow}]-\mu\:(n_{\downarrow} +
n_{\uparrow})$.
(We neglect non-conservation of spin due to surface terms.) 
In the spirit of the Landau theory of critical phenomena we
minimize this functional in the vicinity of the ``critical'' point
$\mu_c$, where the cross-over from full to partial polarization
occurs. Near this point we have $n_{\downarrow} \ll n_{\uparrow}$ and
the condition for the minimum of the free energy becomes

\begin{equation} \label{eq:free_energy}
\begin{array}{lclcl}
\frac{\partial F}{\partial n_{\uparrow}} & = &
\alpha + \alpha' \: \delta n_{\uparrow} + \gamma n_{\downarrow}-\mu&=&0 \\[2mm]
\frac{\partial F}{\partial n_{\downarrow}} & = &
\beta + \beta' n_{\downarrow}  + \gamma \: \delta n_{\uparrow}-\mu&=&0,
\end{array}
\end{equation}
where we have made the linearization $n_{\uparrow} =
n_{\uparrow}^0 + \delta n_{\uparrow}$ for the majority spins and
assumed that the leading terms are linear in $\delta n_{\uparrow}$
and $n_{\downarrow}$. The solution for the minority spin density
in the case of $\mu > \mu_c$ is $n_{\downarrow} \propto (\mu -
\mu_c)$ which combined with the 1d property that $n_{\downarrow}^2
\propto \varepsilon_F^{\downarrow} = \Delta$ 
leads to Eq.~(\ref{eq:Deltamu}). In the other
case, $\mu <\mu _{c}$, $\Delta$ is the energy gap for adding a
minority spin and it is caused by the interaction energy. Thus
again within the same simplified approach, we expect $\Delta$ to
be proportional to the density of majority spins, and hence
$\Delta = D(\mu -\mu_{c})$.

An in-plane magnetic field {\bf B} is readily taken into account
by adding Zeeman energy terms and substituting
\begin{equation} \label{eq:zeeman}
\varepsilon_{\uparrow}^s \rightarrow
\varepsilon^s_{\uparrow}-g\: \mu_B |{\bf B}|, \qquad
\varepsilon^s_{\downarrow} \rightarrow
\varepsilon^s_{\downarrow}+g\: \mu_B |{\bf B}|.
\end{equation}

\begin{figure}
\centerline{\epsfysize=40mm\epsfbox{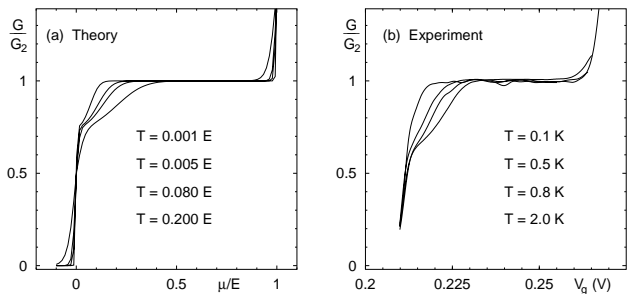}}
\narrowtext
\caption{\label{fig:GT}
(a) The conductance from
Eqs.~(\protect\ref{eq:conductance}) and~(\protect\ref{eq:Deltamu})
with $C=0.5$, $D=1.0$ and $\mu_c=\varepsilon^s_\uparrow$. All energies are
given in units of the transverse mode subband spacing $E$.
(b) Experimental results from
Ref.~\protect\cite{Kristensen98b}.
}
\end{figure}

\begin{figure}
\centerline{\epsfysize=40mm\epsfbox{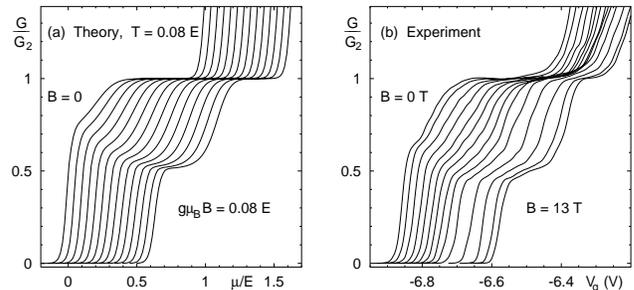}}
\narrowtext
\caption{\label{fig:GB}
(a) The conductance at fixed $T=2.0~E$ as in
Fig.~\protect\ref{fig:GT}(a) with an in-plane magnetic field from
0 to 0.08~$E/g\mu_B$. (b) Experimental results from
Ref.~\protect\cite{Thomas96}. For clarity, the curves are
off-set horizontally.
}
\end{figure}

{\em Experimental implications of the model}. In the following we
discuss how the model can explain the experimental observations
(e1)-(e6) summarized above. To facilitate comparison with experiment
we have added a spin-degenerate subband with $\varepsilon^s_2 = \varepsilon^s_\uparrow
+E$, where $E$ is a constant transverse-mode subband-spacing. 
In Fig.~\ref{fig:GT} observations (e1) and (e2) are clearly seen in
the model calculation. The plateau-like feature in the figure is due to the
specific functional form of $\Delta(\mu)$ in Eq.~(\ref{eq:Deltamu})
which as mentioned before ensures the fulfilment of condition 
Eq.~(\ref{eq:condition}). In this idealized case with a
step-function transmission coefficient the plateau appears
at $0.75$ as discussed above. 
Observation (e3) follows trivially from Eq.~(\ref{eq:dIdmu}) with
the activation temperature $T_a = \Delta(\mu)$. Assuming that in
the vicinity of $\mu_c$ the chemical potential depends linearly on
the gate voltage $V_g$ Eq.~(\ref{eq:Deltamu}) immediately predicts
(e4) with the exponent $\alpha=2$. We now turn to the
characteristic magnetic field dependence (e5) of the 0.7 plateau
at a fixed temperature. The result of the model calculation using
Eqs.~(\ref{eq:conductance}), (\ref{eq:Deltamu}) and~(\ref{eq:zeeman}) is
shown in Fig.~\ref{fig:GB}. In accordance with observation the 0.7
anomaly develops smoothly into an ordinary Zeeman split 0.5
plateau. The last experimental observation (e6) concerns finite
bias. This brings us into a strong non-equilibrium situation which
is outside the scope of the present work. However, considering a
small finite bias not too far from the equilibrium case, we
do find that the 0.75 plateau rises, which gives
additional support for the picture presented here.

{\em Non-ideal transmission}. Our idealized model with a
step-function transmission coefficient predicts an
anomaly around 0.75 rather than around 0.6 - 0.7 as usually
observed in the experiments (see Figs.~\ref{fig:GT}
and~\ref{fig:GB}). When we include more realistic transmission
coefficients, ${\cal T}_{\sigma}(\varepsilon)$, allowing for
resonances to occur this discrepancy in fact
finds a natural explanation. In accordance with the LB formalism
we replace Eq.~(\ref{eq:dIdmu}) by
\begin{equation} \label{eq:dIdmuT}
G(T)  = \frac{1}{2} G_2
\sum_{\sigma=\uparrow,\downarrow}
\int_{-\infty}^{\infty} d\varepsilon \:
{\cal T}_{\sigma}(\varepsilon) \: f'[\varepsilon-\mu].
\end{equation}
\begin{figure}
\centerline{\epsfysize=40mm\epsfbox{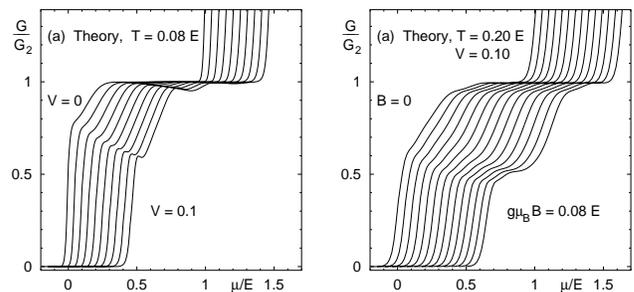}}
\narrowtext
\caption{ \label{fig:Gbscat}
(a) The effect of increasing backscattering on the conductance ranging
from ideal transmission ($V$=0, as in Fig~\protect\ref{fig:GT}(a))
to strong non-ideal transmission ($V$=0.1) governed  by the parameter
$V$ of the transmission coefficient\protect\cite{TofE}. (b) The change
of Fig.~\protect\ref{fig:GB}(a) due to non-ideal transmission.
}
\end{figure}
In contrast to the idealized model this expression is not
universal, but two general features can be expected. First of
all, due to the conditions Eqs.~(\ref{eq:condition})
and~(\ref{eq:Deltamu}) the quasi-plateau 
persists. Secondly, mainly the transmission coefficient 
of minority spin band will be affected which results in a suppression
of the anomalous plateau while the integer plateau remains close
to 1. In Fig.~\ref{fig:Gbscat} this is illustrated by using the
transmission coefficient for a rectangular potential
barrier\cite{TofE}. This choice of ${\cal T}_\sigma(\varepsilon)$ might be
particularly relevant to the recent experiments on long quantum
wires\cite{Reilly00}. 

{\em Suppression of shot noise.} Deeper insight in the nature of the
0.7 anomaly may be obtained from shot noise measurements. Below we
contrast the standard LB treatment \cite{Lesovik89,Buttiker90}
with our model. In the standard spin degenerate case the conductance
is interpreted in terms of 
an overall reduction of the transmission coefficient 
${\cal T}_0$ and the noise spectrum at the 0.7 anomaly is
\begin{equation} \label{eq:LBnoise}
\langle I_{\omega} I_{-\omega} \rangle_{\omega\rightarrow 0}= e
(1-{\cal T}_0) I= G_2 {\cal T}_0 (1-{\cal T}_0) \Delta \mu
\end{equation}
with ${\cal T}_0=0.7$. In our model the 0.7 anomaly comes from
thermal depopulation of spin subbands and not from a reduced
transmission, and the noise spectrum is
\begin{equation} \label{eq:BCFnoise}
\langle I_{\omega} I_{-\omega}\rangle_{\omega\rightarrow 0} = 
\frac{1}{2} G_2[(1-{\cal T}_\uparrow) {\cal T}_\uparrow+ 
(1-{\cal T}_\downarrow){\cal T}_\downarrow] \Delta \mu,
\end{equation}
which in the simple version with ${\cal T}_\sigma(\varepsilon) =
\Theta(\varepsilon-\varepsilon^s_\sigma)$ leads to a vanishing shot noise.
When non-ideal transmission is included, our model does not
predict a universal noise contribution for the minority spins. 
We can, however, see that while the 0.7 quasi-plateau may 
be strongly reduced by additional backscattering (${\cal T}_\downarrow \ll
1$), the transmission in the majority spin subband remains large
(${\cal T}_\uparrow \approx 1$), and Eq.~(\ref{eq:BCFnoise}) yields
\begin{equation}
\langle I_{\omega} I_{-\omega} \rangle_{\omega\rightarrow 0}
\ll G_2 {\cal T}_0 (1-{\cal T}_0) \Delta\mu.
\end{equation}
Thus in general our model predicts a strong suppression of
shot noise as compared to the standard result Eq.~(\ref{eq:LBnoise}).
This effect may already have been observed (see Fig.~3 in
Ref.~\cite{Liu98}). 

In summary, we have presented a phenomenological model which can
account for the experimental observations of the anomalous 0.7
conductance plateau in mesoscopic QPCs. The model is built on an
assumption of an effective instantaneous partial polarization seen by
the transversing electrons, while the ground state itself needs not
have a finite magnetic moment. We hope that the present picture can
inspire future work on microscopic theories of enhanced spin
correlations in open mesoscopic systems.

{\em Acknowledgements.} We are grateful for the experimental data
provided by Anders Kristensen and James Nicholls.
H.B.\ and V.V.C.\ both acknowledge support from the Danish Natural
Science Research Council through Ole R{\o}mer Grant No.\ 9600548.


\end{multicols}

\begin{references}


\bibitem{vanWees88} B.J. van Wees {\it et al.},
  Phys.Rev.Lett. {\bf B 60}, 848 (1988).

\bibitem{Wharam88} D.A. Wharam {\it et al.}, 
  J.Phys.C {\bf 21}, L209 (1988).

\bibitem{vanHouten92} for a review see {\it e.g.\/}
  H. Van Houten, C.W.J. Beenakker, and B. van Wees,
  p. 9 in {\em Nanostructured Systems, M. Reed vol.\ ed.},
  {\em Semiconductors and Semimetals {\bf 35}, 
  R.K. Williamson, A.C. Beer and R. Weber eds.},
  Academic Press, 1992.

\bibitem{Thomas96} K.J. Thomas {\it et al.}, 
  Phys. Rev. Lett. {\bf 77}, 135 (1996).

\bibitem{Thomas98} K.J. Thomas {\it et al.},
  Phys. Rev. B {\bf 58}, 4846 (1998).

\bibitem{Liang99} C.-T. Liang {\it et al.\/}, 
  cond-mat/9907379.

\bibitem{Kristensen98a} A. Kristensen {\it et al.}, 
  J. Appl. Phys. {\bf 83}, 607 (1998).

\bibitem{Kristensen98b} A. Kristensen {\it et al.},
  Physica B {\bf 249-251}, 180 (1998).

\bibitem{Kristensen98c} A. Kristensen {\it et al.}, 
  cond-mat/9808007.

\bibitem{Reilly00} D. J. Reilly {\it et al.}, 
  cond-mat/0001174.

\bibitem{Berggren} C.-K. Wang and K.-F. Berggren,
  Phys. Rev. B {\bf 57}, 4552 (1998).

\bibitem{Spivak} B. Spivak and F. Zhou,
  cond-mat/9911175.

\bibitem{Lieb} E. Lieb and D. Mattis,
  Phys. Rev. {\bf 125}, 164 (1962).

\bibitem{Mermin} N.D. Mermin and H. Wagner,
  Phys. Rev. Lett. {\bf 17}, 1133 (1966).

\bibitem{Tarucha95} S. Tarucha, T. Honda, and T.Saku,
  Solid State Commun. {\bf 94}, 413 (1995).

\bibitem{Yacobi96} A. Yacobi {\it et al.}, 
  Phys. Rev. Lett. {\bf 77}, 4612 (1996).

\bibitem{Tscheuschner96} R.D. Tscheuschner and A.D. Wieck,
  Superlattices and Microstructures {\bf 20} 615 (1996).

\bibitem{Alekseev98} A.Y. Alekseev and V.V. Cheianov, 
  Phys. Rev. B {\bf 57}, 6834 (1998).

\bibitem{Kane92} C.L. Kane and M.P.A. Fisher,
  Phys. Rev. B {\bf 46}, 15233 (1992).

\bibitem{Maslov} D.L. Maslov, 
  Phys. Rev. B {\bf 52}, R14368 (1995).

\bibitem{Maslov95} D.L. Maslov and M. Stone,
  Phys. Rev. B {\bf 52}, R5539 (1995).

\bibitem{Ponomarenko95} V.V. Ponomarenko,
  Phys. Rev. B {\bf 52}, 8666 (1995).

\bibitem{Safi95} I. Safi and H.J. Schultz,
  Phys. Rev. B {\bf 52}, 17040 (1995).

\bibitem{Bruus} H. Bruus and K. Flensberg,
  Semicond. Sci. Technol. {\bf 13}, A30 (1998).



\bibitem{TofE} The transmission coefficient $T(\varepsilon)$ for a
rectangular potential barrier of height $V$ and length $a$ is given by
$T(\varepsilon) =  1/|\cosh(2\kappa a) + i \gamma \sinh(2\kappa
a)|^2$, where $k = \sqrt{\varepsilon}$, $\kappa=\sqrt{V-\varepsilon}$,
and $\gamma = (\kappa/k + k/\kappa)/2$. We set $a=5$.

\bibitem{Lesovik89} G.B. Lesovik, 
  Pis'ma Zh. Eksp. Theo. Fiz. {\bf 49}, 513 (1989)
  [Sov. Phys. JETP Lett. {\bf 49}, 592 (1989)].

\bibitem{Buttiker90} M. B\"{u}ttiker, 
  Phys. Rev. Lett. {\bf 65}, 2901 (1990).

\bibitem{Liu98} R.C. Liu, B. Odom, Y. Yamamoto, and S. Tarucha, Nature
{\bf 391}, 263 (1998), and private communication.

\end{references}
\end{document}